\documentclass[12pt,preprint]{aastex}

\begin{document}
\title{Observation of VHE Gamma Radiation from HESS J1834-087/W41 with
the MAGIC Telescope}

\author{
 J.~Albert\altaffilmark{a}, 
 E.~Aliu\altaffilmark{b}, 
 H.~Anderhub\altaffilmark{c}, 
 P.~Antoranz\altaffilmark{d}, 
 A.~Armada\altaffilmark{b}, 
 M.~Asensio\altaffilmark{d}, 
 C.~Baixeras\altaffilmark{e}, 
 J.~A.~Barrio\altaffilmark{d}, 
 M.~Bartelt\altaffilmark{f}, 
 H.~Bartko\altaffilmark{g,*}
% ...
}
 \altaffiltext{a} {Universit\"at W\"urzburg, D-97074 W\"urzburg, Germany}
 \altaffiltext{b} {Institut de F\'\i sica d'Altes Energies, Edifici Cn., E-08193 Bellaterra (Barcelona), Spain}
 \altaffiltext{c} {ETH Zurich, CH-8093 Switzerland}
 \altaffiltext{d} {Universidad Complutense, E-28040 Madrid, Spain}
 \altaffiltext{e} {Universitat Aut\`onoma de Barcelona, E-08193 Bellaterra, Spain}
 \altaffiltext{f} {Universit\"at Dortmund, D-44227 Dortmund, Germany}
 \altaffiltext{g} {Max-Planck-Institut f\"ur Physik, D-80805 M\"unchen, Germany}
% ...
\altaffiltext{*} {correspondence: H. Bartko, hbartko@mppmu.mpg.de}


\begin{abstract}

Recently, the HESS array has reported the detection of
$\gamma$-ray emission above a few hundred GeV from eight new
sources located close to the Galactic Plane. The source HESS
J1834-087 is spatially coincident with SNR G23.3-0.3 (W41). Here
we present MAGIC observations of this source, resulting in the
detection of a differential $\gamma$-ray flux consistent with a
power law, described as $\mathrm{d}N_{\gamma}/(\mathrm{d}A
\mathrm{d}t \mathrm{d}E) = (3.7 \pm 0.6) \times 10^{-12}
(E/\mathrm{TeV})^{-2.5 \pm 0.2} \ \mathrm{cm}^{-2}\mathrm{s}^{-1}
\mathrm{TeV}^{-1}$. We confirm the extended character of this
flux. We briefly discuss the observational technique used, the
procedure implemented for the data analysis, and put this
detection in the perspective of the molecular environment found in
the region of W41. We present $^{13}$CO and $^{12}$CO emission
maps showing the existence of a massive molecular cloud in spatial
superposition with the MAGIC detection.

\end{abstract}


\section{Introduction}

In the Galactic Plane scan performed by the HESS Cherenkov array
in 2004, with a flux sensitivity of $3\%$ Crab units for
$\gamma$-rays above $200$ GeV, eight sources were discovered
\citep{Aharonian2005a,Aharonian2005b}. One of the newly detected
$\gamma$-ray sources is HESS J1834-087 
which is found to be, in projection, spatially coincident with SNR
G23.3-0.3 (W41). The possibility of a random correlation between
the VHE source and SNR G23.3-0.3 was estimated to be 6\% for the
central region of the Galaxy \citep{Aharonian2005a}. The high
energy source could also be connected to the old pulsar
PSRJ1833-0827 (Gaensler \& Johnston 1995), which would be
energetic enough as to power HESS J1834-087. However, its location at
24 arc minutes away from the center of HESS J1834-087, renders an
association unlikely \citep{Aharonian2005a,Aharonian2005b}. In
addition, there is also no extended PWN detected so far, whereas
HESS J1834-087 has been found to be extended: A brightness distribution
$\rho \sim \exp(-r^2/2\sigma^2)$ with a size $\sigma=(0.09 \pm
0.02)^{\circ}$ has been reported by HESS \citep{Aharonian2005b}.

Here, we present observations of HESS J1834-087 with the Major Atmospheric
Gamma Imaging Cherenkov telescope (MAGIC). We briefly
discuss the observational technique used and the procedure
implemented for the data analysis, derive a very high energy
$\gamma$-ray spectrum of the source, and analyze it in comparison
with other observations, including the molecular environment found
in the region of W41.

\section{Observations}

MAGIC (see e.g., \citet{MAGIC-commissioning,CortinaICRC} for a
detailed description) is the largest single dish Imaging Air
Cherenkov Telescope (IACT) in operation. Located on the Canary
Island La Palma ($28.8^\circ$N, $17.8^\circ$W, 2200~m a.s.l.), the
telescope has a 17-m diameter tessellated parabolic mirror,
supported by a light weight space frame of carbon fiber reinforced
plastic tubes. It is equipped with a 576-pixel $3.5^\circ$
field-of-view enhanced quantum efficiency photomultiplier (PMT)
camera. The analogue PMT signals are transported via optical
fibers to the trigger electronics
and are read out by a 300 MSamples/s FADC system.

At La Palma, HESS J1834-087 culminates at about $37^\circ$ zenith angle
(ZA). This ZA increases the energy threshold for MAGIC
observations, but also, it provides a larger effective collection
area. The sky region 
around the location of HESS J1834-087 has a relatively high and
non-uniform level of background light. Within a distance of
1$^{\circ}$ from HESS J1834-087, there are 3 stars brighter than
8$^{\mathrm{th}}$ magnitude, with the star field being brighter in
the region NW of the source.
MAGIC observations were carried out in the false-source tracking
(wobble) mode \citep{wobble}. The sky directions (W1, W2) to be
tracked are such that in the camera the sky brightness distribution
relative to W1 is similar to the one relative to W2. The source
direction is in both cases $0.4^\circ$ offset from the camera
center.  These two tracking positions are shown by white stars in
figure \ref{fig:disp_map}. For each
tracking position two background control regions are used, which
are located symmetrically to the source region (denoted by the central
white circle) with respect to the
camera center. During wobble mode data taking, 50\% of
the data is taken at W1 and 50\% at W2, switching (wobbling)
between the 2 directions every 30 minutes. This observation mode
allowed a reliable background estimation least affected by the
medium-scale ZA and the inhomogeneous 
star field. HESS J1834-087 was observed for a total of 20 hours in
the period August-September 2005 (ZA $\leq 45^\circ$). In total,
about 12 million triggers have been recorded.

\section{Data Analysis}

The data analysis was carried out using the standard MAGIC
analysis and reconstruction software \citep{Magic-software}, the
first step of which involves the calibration of the raw data
\citep{MAGIC_calibration}. It follows the general steps presented
in \citep{MAGIC_1813,MAGIC_GC}: After calibration, image cleaning
tail cuts of 10 photoelectrons (ph.~el.) for image core pixels and
5 ph.~el. (boundary
pixels) have been applied (see e.g. \citep{Fegan1997}). These tail
cuts are accordingly scaled for the larger size of the outer
pixels of the MAGIC camera. The camera images are parameterized by
image parameters \citep{Hillas_parameters}. In this analysis, the
Random Forest method (see \citet{RF,Breiman2001} for a detailed
description) was applied for the $\gamma$/hadron separation (for a
review see e.g. \citet{Fegan1997}) and the energy estimation.
For the training of the Random Forest a sample of Monte Carlo (MC)
generated $\gamma$-showers \citep{Majumdar2005} was used together
with about 1\% randomly selected events from the measured wobble
data. The MC $\gamma$-showers were generated between 35$^\circ$
and 45$^\circ$ ZA with energies between 10~GeV and 30~TeV with a
SIZE distribution equal to the one of the selected data events
for the training. The
source-position independent image parameters  SIZE, WIDTH, LENGTH,
CONC \citep{Hillas_parameters} and the third moment of the ph.~el.
distribution along the major image axis were selected to
parameterize the shower images. After the training, the Random
Forest method allows to calculate for every event a parameter,
so-called hadronness, which is a measure of the probability that
the event is not $\gamma$-like. The $\gamma$-sample is defined by
selecting showers with a hadronness below a specified value. An
independent sample of MC $\gamma$-showers was used to determine
the cut efficiency. 

The analysis at similar ZA angles was developed and verified using
Crab nebula data taken in September 2005, see also
\citep{MAGIC_1959}. The Crab energy spectrum, as determined by our
studies, was consistent with measurements from other experiments
(see Fig. \ref{fig:spectrum}, dot-dashed line).

For each event the arrival direction of the primary $\gamma$-ray
candidate in sky coordinates is estimated using the DISP-method
\citep{wobble,Lessard2001,MAGIC_disp}. A conservative lower SIZE
cut of 200 ph.~el. is applied to select a subset of events with
superior angular resolution. The corresponding analysis energy
threshold is about 250~GeV.

\begin{figure}[t]
\begin{center}
\includegraphics[totalheight=7cm]{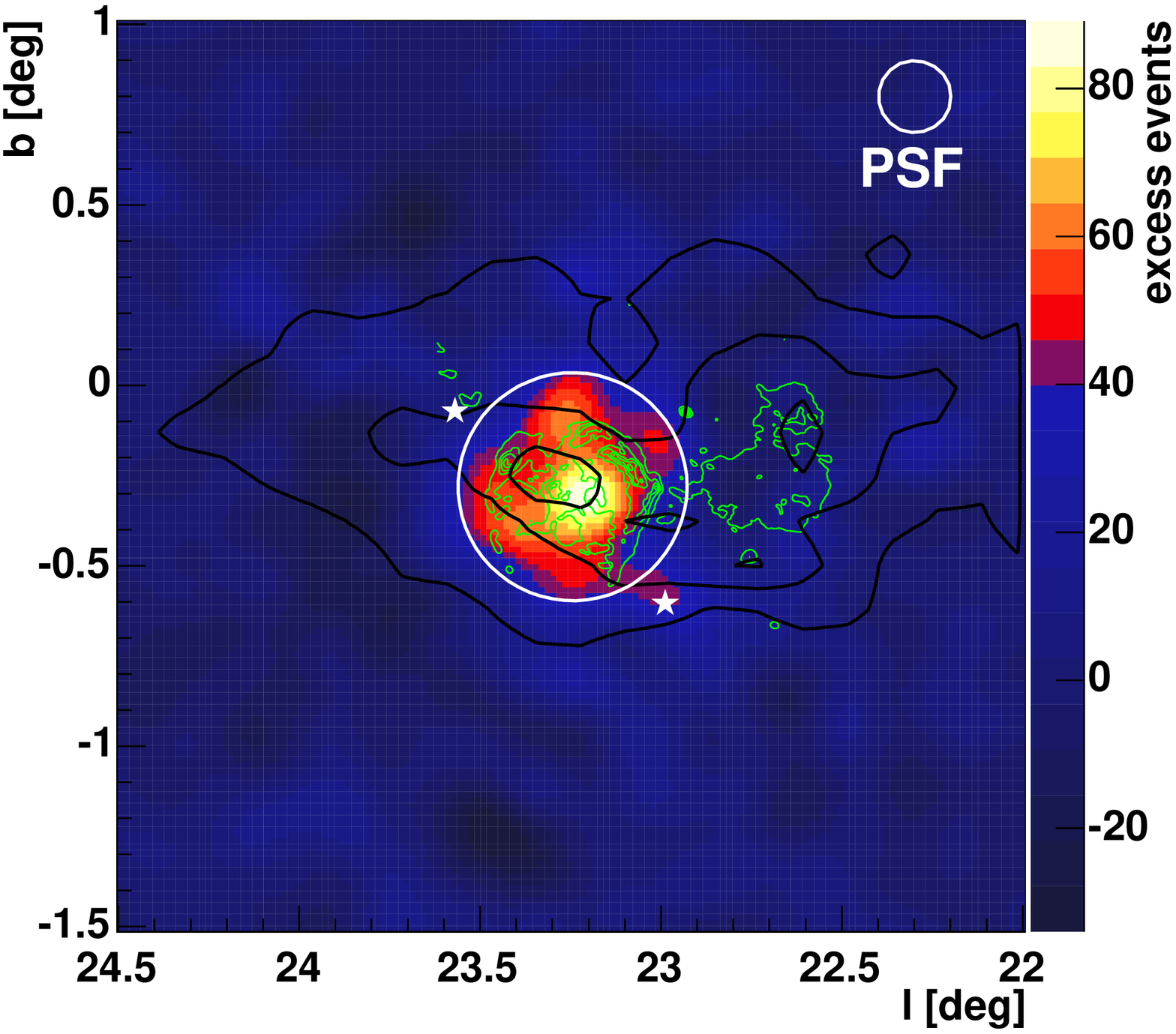}
\end{center}
\caption{Sky map of $\gamma$-ray candidate events (background
subtracted) in the directions of HESS J1834-087 for an energy threshold
of about 250~GeV. The source is clearly extended with respect to
the MAGIC PSF. The two white stars denote the tracking positions
in the wobble mode. Overlayed are $^{12}$CO  emission contours
(black) from \citet{Dame2001} and contours of 90 cm VLA radio data
from \citet{White2005} (green). The $^{12}$CO contours are at
25/50/75 K km/s, integrated from 70 to 85 km/s in velocity, the
range that best defines the molecular cloud associated with W41.
The contours of the radio emission are at
0.04/0.19/0.34/0.49/0.64/0.79 Jy/beam, chosen for best showing
both SNRs G22.7-0.2 and G23.3-0.3 at the same time. Clearly, there
is no superposition with SNR G22.7-0.2. The central white circle
denotes the source region integrated for the spectral analysis.}
\label{fig:disp_map}
\end{figure}

Figure \ref{fig:disp_map} shows the sky map of $\gamma$-ray
candidates (background subtracted, see e.g. \citep{Rowell2003})
from the direction of HESS J1834-087. It is folded with a
two-dimensional Gaussian with a standard deviation of
$0.072^{\circ}$ and a maximum of one.
The MAGIC $\gamma$-ray PSF (standard deviation of a two dimensional 
Gaussian fit to the non-folded brightness profile of a point source) 
is $0.1\pm0.01^{\circ}$. The folding of the sky map serves
to increase the signal-to-noise ratio by smoothing out
statistical fluctuations. However, it somewhat degrades
the spatial resolution.
The sky map is overlayed with contours of 90 cm
VLA radio data (green) from \citet{White2005} (20 cm radio data
from the same reference are overlayed in the following figures)
and $^{12}$CO emission contours from
\citet{Dame2001} (black), integrated in the velocity range 70 to
85 km/s, the range that best defines the molecular cloud
associated with W41. The MAGIC excess is centered at (RA,
DEC)=(18$^\mathrm{h}$34$^\mathrm{m}$27$^\mathrm{s}$,
-8$^\circ$42'40''). The statistical error is 0.5', the systematic 
pointing uncertainty is estimated to be 2' (see
\citet{Bretz2003}). 
A fit of a two dimensional Gaussian brightness
profile to the non-folded sky map yields after subtraction the MAGIC 
$\gamma$-ray PSF inquadrature an intrinsic source extension of 
$\sigma = (0.14 \pm 0.04)^{\circ}$ (the extension reported by 
HESS is $0.09 \pm 0.02)^{\circ}$ \citep{Aharonian2005b}). Both, 
position and extension, coincide well with the shell-type 
SNR G23.3-0.3 (W41).

\begin{figure}[!h]
\begin{center}
\includegraphics[totalheight=5cm]{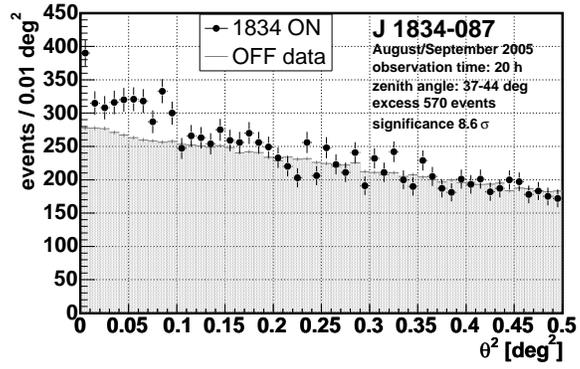}
\end{center}
\caption{Distributions of $\theta^2$ values for the source (full
circles) and background control region (shaded histogram), see
text, for an energy threshold of about 250~GeV. It is seen also
here that the source is clearly extended with respect to the MAGIC
PSF ($\sigma^2 \sim 0.01 \mathrm{deg}^2$).} \label{fig:theta2}
\end{figure}

\begin{figure}[!h]
\begin{center}
\includegraphics[width=\textwidth]{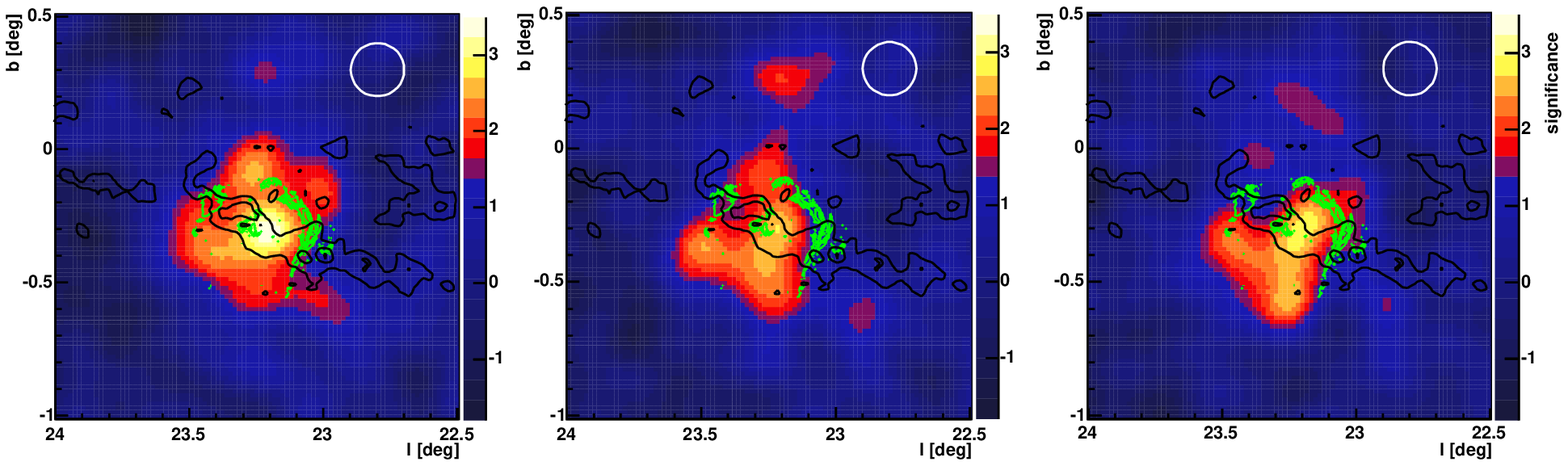}
\end{center}
\caption{Morphology
of HESS J1834-087 above three different lower cuts in SIZE (200, 300,
600 ph.el.), corresponding to energy thresholds of 250, 360, 590
GeV. The color scale shows the excess significance. Overlayed are
contours  of 20 cm VLA radio data from \citet{White2005} (green)
and $^{13}$CO emission contours (black) from \citet{Jackson2006}.
The contours of the radio emission are at 0.0035 Jy/beam. The
$^{13}$CO contours are at 10/20/30 K km/s, integrated from 70 to
85 km/s in velocity, as was the $^{12}$CO data in figure 
\ref{fig:disp_map}. The white circle indicates the MAGIC PSF which 
is about 0.1 deg for all three lower SIZE cuts.} \label{fig:three_disp_maps}
\end{figure}

\begin{figure}[!h]
\begin{center}
\includegraphics[totalheight=5cm]{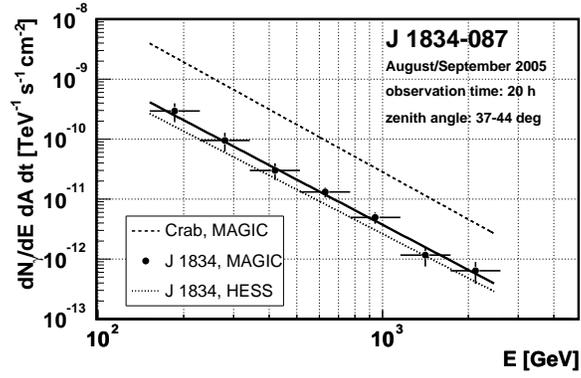}
\end{center}
\caption{VHE $\gamma$-ray spectrum of HESS J1834-087 (statistical errors only).
The solid line shows the result of a power-law fit to the data points. The
dotted line shows the result of the HESS collaboration \citep{Aharonian2005b}.
The dashed line shows the spectrum of the Crab nebula as measured by MAGIC
\citep{Crab_MAGIC}.
} \label{fig:spectrum}
\end{figure}

Figure \ref{fig:theta2} shows the distribution of the squared
angular distance, $\theta^2$, between the reconstructed shower
direction and the excess center. The observed excess in the
direction of HESS J1834-087 has a significance of 8.6$\sigma$ for
$\theta^2 \leq 0.1 \mathrm{deg}^2$. Figure
\ref{fig:three_disp_maps} shows images of HESS J1834-087 with three
different lower cuts on SIZE (200, 300, 600 ph.el), corresponding
to energy thresholds of about 250, 360 and 590 GeV. As in figure
\ref{fig:disp_map} the background subtracted sky maps are folded
with a two-dimensional Gaussian, but here the color scale shows
directly the excess significance. The total observed excess
significance for $\theta^2 \leq 0.1 \mathrm{deg}^2$ (corresponding
to the sky region inside the central white circle of figure
\ref{fig:disp_map}) are 8.6$\sigma$, 7.8$\sigma$ and 7.3$\sigma$
for the three lower cuts on SIZE. Overlayed are contours of 20 cm
VLA radio data from \citet{White2005} (green) and $^{13}$CO
emission contours (black) from \citet{Jackson2006}. The contours
of the radio emission are at 0.0035 Jy/beam. The $^{13}$CO
contours are integrated from 70 to 85 km/s in velocity, as was the 
$^{12}$CO data in figure \ref{fig:disp_map}. For all three SIZE
cuts the MAGIC PSF is about $0.1^{\circ}$, and the source
position, extension and morphology stay roughly constant. The
characteristics of the MAGIC observation are compatible within errors
with the measurement of HESS \citep{Aharonian2005b}.


For the spectral analysis a sky region of maximum angular distance
of $\theta^2 = 0.1 \mathrm{deg}^2$ around the excess center
(indicated by the white circle in Figure \ref{fig:disp_map}) has
been integrated. Figure \ref{fig:spectrum} shows the reconstructed
very high energy $\gamma$-ray spectrum
($\mathrm{dN}_{\gamma}/(\mathrm{dE}_{\gamma} \mathrm{dA}
\mathrm{dt})$ vs. true $\mathrm{E}_{\gamma}$) of HESS J1834-087 after
correcting (unfolding) for the instrumental energy resolution
\citep{Anykeev1991}. The horizontal bars indicate the bin size in
energy, the marker is placed in the bin center on a logarithmic
scale. The full line shows the result of a forward unfolding
procedure: A simple power law spectrum is fitted to the measured
spectrum ($\mathrm{dN}_{\gamma}/(\mathrm{dE}_{\gamma} \mathrm{dA}
\mathrm{dt})$ vs. estimated $\mathrm{E}_{\gamma}$) taking the full
instrumental energy migration (true $\mathrm{E}_{\gamma}$ vs.
estimated $\mathrm{E}_{\gamma}$) into account as described in
\citet{Mizobuchi2005}. The result is given by
($\chi^2/\mathrm{n.d.f}=7.4/7$):
\begin{eqnarray*}
\frac{ \mathrm{d}N_{\gamma}}{\mathrm{d}A \mathrm{d}t \mathrm{d}E}
= (3.7 \pm 0.6) \times  10^{-12}
\left(\frac{E}{\mathrm{TeV}}\right)^{-2.5 \pm 0.2} \nonumber
\\
\hspace{4cm} \mathrm{cm}^{-2}\mathrm{s}^{-1}
\mathrm{TeV}^{-1}.
\end{eqnarray*}
The quoted errors are statistical. The systematic error is
estimated to be 35\% in the flux level determination and 0.2 in
the spectral index, see also \citep{MAGIC_GC}. Within the
observation time (weeks) no flux variations exceeding the
measurement errors have been observed. Also, the flux is
compatible within errors with the measurement of HESS made one
year earlier.

\section{Discussion and concluding remarks}


SNRs as gamma-ray sources have been extensively discussed in the
past (e.g., see Torres et al. 2003 for a review). Due to the
spatial coincidence between the VHE $\gamma$-ray source and the
SNR G23.3-0.3 (W41), this SNR appears to be the natural candidate
for generating the observed $\gamma$-ray emission.
%
%
W41 is an asymmetric shell-type SNR, with a diameter of 27 arc
minutes. It is included in Green's catalog \citep{Green2004},
and has a spectral index of 0.5, and a flux density of 70 Jy at 1
GHz. It was mapped in radio with the VLA array at 330 MHz (Kassim
1992) and at 20 and 90 cm (see White, Becker \& Helfand 2005),
following earlier studies (see, e.g., Ariskin \& Berulis 1970,
Shaver \& Goss 1970, and references therein). It is partially
overlapping with SNR G22.7-0.2 (see, e.g., Fig. 9 of Kassim 1992), although
the latter is not in coincidence with the peak of the very high
energy source (see Figure 1 above). No Chandra and XMM
observations of W41 are publicly available yet.



W41 was associated with a very large molecular complex called
``[23,78]'' in Dame et al. (1986). There, it was concluded that there
are probably two large clouds blended at that position in $l-b-v$
space, one in the near side of the 4 kpc arm and another in the
far side of the Scutum Arm.
The giant molecular cloud associated with W41 is
best defined by integrating the CO emission from
70 to 85 km/s in velocity. The CO emission peaks
near l=$23.3^{\circ}$, b=$-0.3^{\circ}$, v=78 km/s; the near 
kinematic distance of this peak is 4.9 kpc. The
peak is marked by the central black contour in
Figure 1, which lies very close to the VHE
source. The total H$_{2}$ mass of the cloud, computed
over the range l=$22^{\circ}$ to $24.25^{\circ}$, b=$-0.75^{\circ}$ to
$0.5^{\circ}$, and v=70 to 85 km/s, and assuming a
distance of 4.9 kpc, is $2.1 \times 10^6$ M$_{\odot}$.  This mass
is necessarily an upper limit since, as
mentioned above, there is certainly an emission
contribution from unrelated gas at the far
kinematic distance. Still, the CO peak is so
strong and well defined that it most likely
arises from gas primarily at one location, near
the VHE source, rather than being a random blend
of emissions from the near and far distances. The
total H$_2$ mass of the CO emission peak in Figure 1
(computed over the region l=$23.2^{\circ}$ to $23.4^{\circ}$,
b=$-0.35^{\circ}$ to $-0.15^{\circ}$, v=70 to 85 km/s) is 
$8.8 \times 10^4$ M$_{\odot}$. The higher-resolution $^{13}$CO map in 
figure 3, which was derived from the recently completed
Galactic Ring survey (Jackson et al. 2006),
confirms that the VHE source lies toward the
a local enhancement of molecular material, a giant molecular cloud.


At 5 kpc, the luminosity of HESS J1834-087 between 250 GeV and a few
TeV is about $ 5 \times 10^{34}$ erg s$^{-1}$, similar to the
luminosity of HESS J1813-178 \citep{MAGIC_1813} if that source is
considered associated with SNR G12.8-0.0 at a distance of $\sim 4$
kpc. We note, though, that J1813-178 has been found to be nearly
point like whereas in the present case, a significant extension is
observed. The $\gamma$-ray spectrum of HESS J1834-087 is steeper than
the one of J1813-178. 
From the observed $\gamma$-ray luminosity, and assuming an
acceleration efficiency of hadrons in the order of 3\% and a
supernova power of 10$^{51}$ erg,  the required density of matter
in the $\gamma$-ray production region for hadrons to be mainly
responsible of the observed radiation can be estimated from the
formula $L_{\gamma} \sim (\mathrm{velocity\ of\ light}) \times 
(\mathrm{density}) \times (\mathrm{efficiency\ of\ acceleration}) 
\times (\mathrm{supernova\ power}) \times (\mathrm{p-p\ cross\ 
section}) \times (\gamma\ \mathrm{p-p\ inelasticity})$, and it is of 
about $\sim$ 11 cm$^{-3}$ (see Torres et al. 2003 and references therein).
With the extension of HESS J1834-087, and the gas mass found
to be in the innermost contour of the CO map, i.e.,  in close
superposition with the very high energy source, there is enough
mass to generate the high energy radiation hadronically, even if
only part of the gas is interacting with the SNR shock.


All in all, the observation of HESS J1834-087 using the MAGIC Telescope
confirms a new very high-energy extended $\gamma$-ray source in
the Galactic Plane.  A reasonably large data set was collected
from observations at medium-scale zenith angles to infer the
spectrum of this source up to energies of a few TeV. Above 200
GeV, the differential energy spectrum can be fitted with a power
law of slope $\Gamma=-2.5\pm 0.2$. The results of the independent
observations of the HESS and MAGIC telescopes are in agreement
within errors concerning the level of flux, the spectral shape,
the morphology, and the extension of the source. The coincidence
of the VHE $\gamma$-ray source with SNR G23.3-0.3 (W41) poses this
SNR as a natural counterpart, and although the mechanism
responsible for the high energy radiation remains yet to be
clarified, a massive molecular cloud has been identified in
the region.

We would
like to thank the IAC for the excellent working conditions at the
Observatory de los Muchachos in La Palma. The support of the
German BMBF and MPG, the Italian INFN and the Spanish CICYT is
gratefully acknowledged. This work was also supported by ETH
Research Grant TH~34/04~3 and the Polish MNiI Grant 1P03D01028.
This paper is dedicated to the memory of Nicolaj Pavel.

\clearpage

\clearpage

\clearpage

\clearpage

\end{document}